\documentclass[11pt,a4paper]{article}
\usepackage{epsfig}
\def\eq#1{(\ref{#1})}
\def\fig#1{{Fig.~\ref{#1}}}
\def\re#1{{Ref.~\cite{#1}}}
\def\order#1{\mathcal{O}{(#1)}}
\newcommand{\beq}{\begin{equation}}
\newcommand{\eeq}{\end{equation}}
\newcommand{\beqar}[1]{\begin{eqnarray}\label{#1}}
\newcommand{\eeqar}{\end{eqnarray}}

\newcommand{\lag}{\mathcal{L}}
\newcommand{\vac}{|\epsilon_\mathrm{v}|}
\def\npb#1#2#3{    {\it Nucl. Phys. }{\bf B#1} (#2) #3}
\def\npa#1#2#3{    {\it Nucl. Phys. }{\bf A#1} (#2) #3}
\def\plb#1#2#3{    {\it Phys. Lett. }{\bf B#1} (#2) #3} 
\def\prd#1#2#3{    {\it Phys. Rev. }{\bf D#1} (#2) #3}

\begin{document}
\title {{ \bf Classical gluodynamics in curved space--time \\ and the soft Pomeron \\[0.7cm]}}
\author{\bf D.~Kharzeev\thanks{e-mail: kharzeev@bnl.gov}~$\, ^{a,b}$,
\quad  E.~Levin\thanks{e-mail: leving@post.tau.ac.il}~$\, ^c$,
\quad K. Tuchin\thanks{e-mail: tuchin@phys.washington.edu}~$\, ^{a,c,d}$\\[10mm]
{\it\small $^a$ Institute for Theoretical Physics, University of California,}\\
{\it\small Santa Barbara, CA 93106, USA}\\[0.3cm]
{\it\small $^b$ Physics Department, Brookhaven National Laboratory,}\\
{\it\small Upton, NY 11973-5000, USA\footnote{Permanent address}}\\[0.3cm]
{\it\small $^c$HEP Department, School of Physics and Astronomy,}\\
{\it\small Tel-Aviv University, Ramat Aviv, 69978, Israel}\\[0.3cm]
{\it\small $^d$Institute for Nuclear Theory, 
University of Washington,}\\
{\it\small Box 351550, Seattle, WA, 98195, USA\footnote{Permanent address}}\\[10mm]}

\date{April, 2002}
\maketitle 
\thispagestyle{empty}

\begin{abstract} 
QCD at the classical level possesses scale invariance  
 which is broken by quantum effects. This ``dimensional transmutation'' 
phenomenon can be mathematically described by
 formulating classical gluodynamics  in a curved, conformally flat, 
space--time with non--vanishing cosmological constant. We study QCD 
high--energy scattering in this theory. We find that the properties of 
the scattering amplitude at small momentum transfer are determined by 
the energy density of vacuum fluctuations. The approach gives rise to
the power growth of  the total hadron--hadron cross section with
energy, i.e.\ the Pomeron. The intercept of the Pomeron and the multiplicity 
of produced particles are evaluated. We also speculate 
about a possible link between conformal anomaly and TeV scale quantum gravity. 

\end{abstract}

\begin{flushright}
\vspace{-20cm}
TAUP-2688-2001\\
BNL-NT-02/3 \\
INT--PUB-02-31\\
NSF-ITP-02-34
\end{flushright}

\newpage
\setcounter{page}{1}

\section{Introduction}

At the classical level, QCD is a scale invariant theory. 
Quantum fluctuations, however, give rise to scale anomaly \cite{SA} which is associated with
dimensionful scale $\Lambda\approx 0.2$~GeV. The perturbative
expansion is possible if the typical distance $Q^{-1}$ at which
the interaction occurs is much smaller than $\Lambda^{-1}$, so 
the coupling is weak \cite{GW}: $\alpha_s(Q) \ll 1$.
However, this
does not mean that at short distances non-perturbative contributions can always be neglected.
Indeed, the strength of the semi-classical vacuum fluctuations 
of the gluon field is inversely proportional to the coupling constant and
thus can be large at short distances, when $\sim 1/\alpha_s(Q) \gg 1$. 
The goal of the present work is to derive an effective lagrangian for gluodynamics which 
makes it possible to systematically treat such contributions to the scattering 
amplitudes. 

Gluodynamics is defined by the lagrangian 
\beq\label{LGD}
\lag_\mathrm{gluon}=-\frac{1}{4}F^a_{\mu\nu}F^a_{\mu\nu}\quad.
\eeq
If treated classically, this theory is invariant under the 
scale transformations $ x \rightarrow \lambda x$. Transformation of scale is 
associated with the scale current $s_{\mu}$; 
its divergence is equal to the trace of the energy--momentum tensor:
\beq 
\partial^{\mu} s_{\mu} = \theta_{\mu}^{\mu}(x).
\eeq
Without quantum effects, $\theta_{\mu}^{\mu}(x) = 0$, and the theory is scale invariant. 
Quantum effects break scale invariance \cite{SA}. However, the broken symmetry still 
manifests itself in the following set of 
low energy theorems for different Green functions involving operator
$\theta_{\mu}^{\mu}(x)$ which can be proven by using renormalization group arguments \cite{NSVZ}:  
\beq\label{LET}
i^n\int dx_1\ldots dx_n\langle0|T\{\theta_{\mu_1}^{\mu_1}(x_1),
\ldots,\theta_{\mu_n}^{\mu_n}(x_n)\}|0\rangle_\mathrm{connected}=
\langle\theta_{\mu}^{\mu}(x)\rangle_\mathrm{vac}(-4)^n
\eeq

In \re{MISHI} it was shown that these 
low--energy theorems entirely determine the form of the 
effective low energy lagrangian for gluodynamics, 
which describes semi-classical vacuum
fluctuations of gluon field at large distances. 
It is formulated in terms of an effective scalar dilaton field:
\beq\label{LDIL}
\lag_\mathrm{dilaton}=\frac{\vac}{m^2}\frac{1}{2}e^{\chi/2}
(\partial_{\mu}\chi)^2
+\vac e^\chi(1-\chi)\quad,
\eeq 
where the real scalar field $\chi$ of mass $m$ represents the dilaton,
and $-\vac$ is the vacuum energy density.  

The idea underlying the derivation can be 
formulated in the way which illustrates the analogy between gluodynamics and general relativity: 
QCD coupled to the conformally flat gravity is scale and
conformally invariant in any number of dimensions, and thus such theory is
anomaly--free. However, the vacuum energy density of flat--space gluodynamics
$-\vac$ (which plays a role of cosmological constant in general relativity) 
manifestly breaks scale and conformal symmetry of gluodynamics in 
conformally flat space-time. The low energy effective lagrangian
\eq{LDIL} thus can be derived from the Einstein-Hilbert lagrangian for the
described theory. Note that in the limit $\vac \rightarrow 0$ 
lagrangian \eq{LDIL} is
scale  and conformally invariant, since another dimensionful parameter
$m$  just sets the normalization of the dilaton field $\chi$.
At small dilaton momenta $p\ll m$ the lagrangian \eq{LDIL} satisfies the low
energy theorems \eq{LET}.
It is remarkable that the effective lagrangian involves only one scalar
field. The reason is that the gravity is excited by the trace of energy
momentum-tensor, so there is only one independent Einstein equation.

In the present work we are going to extend the range of validity of
the effective lagrangian \eq{LDIL}
to shorter distances by including the interactions of dilaton field with 
gluons. 
Generally, it can be obtained from the lagrangian of gluodynamics
 by integrating
out soft gluon modes up to the scale $m$ which
thus becomes the dimensionful scale of the theory. This implies that the scale
$\Lambda$ and the dilaton mass $m$ are intimately connected. Actually, as we will show, it appears 
more convenient to construct the generalized effective lagrangian using
Einstein equations once again. In the next section  we employ this method
to obtain  the lagrangian \eq{LAGR}. One can
easily check that it is non-renormalizable. Therefore, we need to specify
the additional dimensionful scale -- the ultra-violet cutoff $M_0$.
Obviously, $M_0$ is a function of $\vac$ and $m$.              
It can be calculated using the requirement of vacuum stability, i.e.\  all
radiative corrections must cancel out to give correct value of vacuum
energy density. Its numerical value turns out to be quite large, 
$M_0\sim 3$~GeV. This is in agreement with \re{NSVZ,FK,SHUR} where it 
has 
been argued that the typical scale of the vacuum gluon density due to
semi-classical fluctuations can indeed be that large. In sec.~2 we
discuss the effective lagrangian \eq{LAGR} in detail and calculate the
scale $M_0$.

In sec.~3 we apply the effective lagrangian \eq{LAGR} to calculate
a contribution of the non-perturbative semi-classical fluctuations of the
QCD vacuum to the pomeron intercept. Experimental data on different reactions
at high energies and small momentum transfer support existence of the pole
in the scattering amplitude responsible for high energy behavior of the
total cross sections \cite{PDBC}. This leads to the power-like behavior 
of
the total cross section
$\sigma_\mathrm{tot}=\varsigma\, s^{\alpha_P-1}$,  
where $\varsigma$ depends upon particular reaction and $\alpha_P$
is the universal parameter called the intercept. After decades of hard
work, understanding the nature of the  
object exchanged  in $t$-channel (pomeron) is still a challenging problem. It turned out that the 
perturbative QCD is far from giving a clear answer even if the scales
inherent to colliding particles are hard. Partly, this is due to diffusion
of the transverse momenta in the partonic cascade towards the
non-perturbative scales. This implies  that the
main contribution to the pomeron stems, perhaps, from the 
non-perturbative sector of QCD. Assuming that some non-perturbative
fluctuations of vacuum dominate the scattering amplitude at high energies
one has to explain the power-like behavior  of the total cross section
$\sigma_\mathrm{tot}\sim s^{\alpha_P-1}$ and to obtain the  
 correct numerical value of $\alpha_P$. 
 A few authors have reported on this
problem. Contribution of instantons to the scattering amplitude at high
energies was investigated in \re{KKL,DIL}. In \re{JANIK} it has been argued
that  
fluctuations of strings in the framework of AdS/CFT theory can
yield the desired result. In Refs.~\cite{TAN,KAD} the spectrum of  
glueballs in QCD was discussed as well as  its relation to the structure 
of the pomeron.
Possible contributions of extra compact
dimensions was discussed in  \re{NUS,KT}. A more phenomenological approach 
has been developed in framework of the BFKL gluon emission with additional 
assumption for such an emission in the non-perturbative QCD domain 
\cite{NN}.
In \re{KHALE} two of us have
used the fact that the non-perturbative QCD vacuum is 
dominated by the  semi-classical fluctuations of the gluon field which is
inversely proportional to the coupling $F\sim 1/g$. The importance of such fluctuations 
can be understood in the following way. Consider a gluon
ladder in each rung of which two gluons in a singlet state are produced
in $s$-channel. Each rung gives contribution of the order $g^2 F^2\sim
\order{1}$ since  the two-gluon singlet state gives contribution
proportional to the trace of energy-momentum tensor. Such ladder diagrams
can be summed up yielding a reasonable
value of the pomeron intercept. In the present paper we suggest a
method for systematic calculation of processes involving such 
non-perturbative modes of gluodynamics.  In particular, we will calculate
the total cross section at high energies, reproduce its
power-like behavior and numerical value of the intercept. 
We also calculate the final state multiplicity in this approach.

We conclude the paper by presenting some speculative ideas on the possible link 
between QCD and gravity.   
Recall that the effective lagrangian \eq{LAGR} can be
 derived from the General Relativity. Thus the gravitational
constant can be expressed in terms of the dilaton mass and the vacuum energy
density, Eq.~\eq{MASS}. Its numerical value  $G=0.2\,\mathrm{GeV}^{-2}$, of course,  
has nothing to do with the Planck gravitational constant. Nevertheless,
it may be possible to interpret this result as a signature of the
strong gravitational interactions at scale much lower than the Planck one.  
If we accept this interpretation, then 
it turns out that the  gravitational effects are essential already at
the GeV scale, in order to be responsible for the QCD vacuum structure.
Several authors have suggested mechanisms by which gravitational
interactions can become strong at scales much larger than the Planck one
\cite{GRAV}. Recently it has been suggested that quantum gravity can
become strong at the TeV scale \cite{DIM}. In sec.~4 we briefly discuss those 
ideas. 

\section{Dilaton effective lagrangian} 
The dilaton effective lagrangian in gluodynamics describing the
propagation and interactions of gluon and dilaton fields can be constructed as a 
generalization of Eqs.~\eq{LDIL},\eq{LGD}. It reads
\beq\label{LAGR}
\lag=\frac{\vac}{m^2}\frac{1}{2}e^{\chi/2} (\partial_{\mu}\chi)^2
+\vac e^\chi(1-\chi)-e^{\chi}(1-\chi)\frac{1}{4}
F^a_{\mu\nu}F^{a\mu\nu}\quad.
\eeq  
Eq.~\eq{LAGR} can be obtained utilizing Legendre transformation of the
lagrangian describing coupling of gluodynamics to gravity
\beq\label{HLA}
\lag'=\left(\frac{1}{8\pi G} R -\frac{1}{4}F^a_{\mu\nu}F^{a\mu\nu}+\vac
\right) \sqrt{-\det g_{\mu\nu}}\quad.
\eeq
The first term in (\ref{HLA}) corresponds to the gravitational
lagrangian $\lag_G$. It is expressed through the Ricci
scalar $R$ which can be easily calculated for the conformally trivial
metric  $g_{\mu\nu}=\eta_{\mu\nu}e^{h(x)}$  
\beq\label{GRAV}
\lag_G=\frac{1}{8\pi G} R\sqrt{-\det g_{\mu\nu}}= 
\frac{1}{8\pi G} 3e^h\left(\partial^2_\mu h+\frac{1}{2} (\partial_\mu h)^2
\right)
=-\frac{1}{8\pi G}\frac{3}{2}e^h (\partial_\mu h)^2
\quad,
\eeq
where we have used integration by parts to derive the last equation
and $\eta_{\mu\nu}=\mathrm{diag}\!\{1,-1,-1,-1\}$. In
(\ref{GRAV}) $G$ is the  gravitational constant. The
gravitational field $h$ couples to the gluodynamics 
by means of the second term  
$$
-\frac{1}{4}F^a_{\mu\nu}F^{a\mu\nu}e^{2h}\quad.
$$
These two terms are invariant under scale and conformal transformations 
in a space-time of arbitrary dimension. However, the scale symmetry in QCD
is    
broken down by non-vanishing gluon condensate $\langle F^a_{\mu\nu}
 F^a_{\mu\nu}\rangle$$>0$. This leads to the scale anomaly which manifests
itself as an anomaly of the QCD energy-momentum tensor  
($\theta^{\mu\nu}$) trace
\beq\label{SPUR}
\theta^\mu_\mu=\frac{\beta(g)}{2g}F^a_{\mu\nu} F^{a\mu\nu}\quad , 
\eeq
where the $\beta$-function 
 in the leading order in the strong coupling $g$ reads  
\beq\label{BETA}
\beta(g)=-\frac{11\, g^3}{(4\pi)^2}\quad.
\eeq
Its vacuum expectation value is 
\beq\label{VAC}
\langle \theta^\mu_\mu\rangle= -4\vac\quad.
\eeq
In order for the energy-momentum tensor, calculated from the lagrangian
(\ref{HLA}) to satisfy \eq{VAC}, the third term in \eq{HLA} which explicitly
breaks the scale invariance has been introduced. This term can also be
thought of as the cosmological constant. In the absense of the
gravitational part \eq{GRAV} of the lagrangian \eq{LAGR} the condition
\eq{VAC} is satisfied trivially at the tree level. Performing the Legendre
transformation of (\ref{HLA}) one arrives at \eq{LAGR} \cite{MISHI}. The
mass of the
dilaton can be expressed through the vacuum energy density and the
gravitational constant as follows
\beq\label{MASS}
m^2=\frac{64\pi}{3}\,\vac\, G\quad.
\eeq
Now we require the energy momentum tensor calculated from the 
lagrangian \eq{LAGR} to satisfy the vacuum normalization condition \eq{VAC}.
This means cancelation of the dilaton and gluon contributions to the
vacuum energy density.  We can  formulate \eq{VAC}
in terms of the vacuum expectation values. To this end let us
calculate the trace of the energy-momentum tensor
\beq\label{TRACE}
\theta_\mu^\mu=\eta^{\mu\nu}\left(
 2\frac{\partial\lag}{\partial\eta^{\mu\nu}}-\eta^{\mu\nu}\lag\right)+
\frac{8\vac}{m^2}\partial_\mu^2\, e^{\chi/2}\quad,
\eeq
where the last term in the right hand side is the 
total derivative \cite{MISHI}.
Equation of motion for the dilaton field is 
\beq\label{EQMOT} 
\frac{\vac}{m^2}\,\partial_\mu\left( e^{\chi/2}\,\partial_\mu\chi\right)-
\frac{\vac}{4m^2}\,e^{\chi/2}\left(\partial_\mu\chi\right)^2+
\chi\, e^\chi\,\vac - 
\chi\, e^\chi\,\frac{1}{4}\,F_{\mu\nu}^aF^{a\mu\nu}=0\quad.
\eeq
Using \eq{EQMOT} in \eq{TRACE} we arrive at
\beq\label{SP}
\theta_\mu^\mu=-4\,\vac\, e^\chi+\chi\, e^\chi\, F_{\mu\nu}^aF^{a\mu\nu}
\quad.
\eeq
 Hence, \eq{VAC} 
is equivalent to the following relation 
\beq\label{NORMREL}
\langle\; \chi \, e^\chi\, F_{\mu\nu}^aF^{a\mu\nu}\;\rangle 
= 4\,\vac\; \langle\,  e^\chi -1\;\rangle\quad.
\eeq
It is clear from  equations \eq{SP} and \eq{EQMOT} that in the  
limit $\vac\rightarrow 0$ the classical equation $\theta_\mu^\mu=0$ 
is recovered. However, quantum corrections can give non--vanishing
contributions to $\langle\theta_\mu^\mu\rangle$ in the same limit
by introducing dimensonal cutoff. Eq.~\eq{NORMREL} assures that this does
not happen in our theory.

\begin{figure}[ht]
\begin{tabular}{ccc}
\epsfig{file=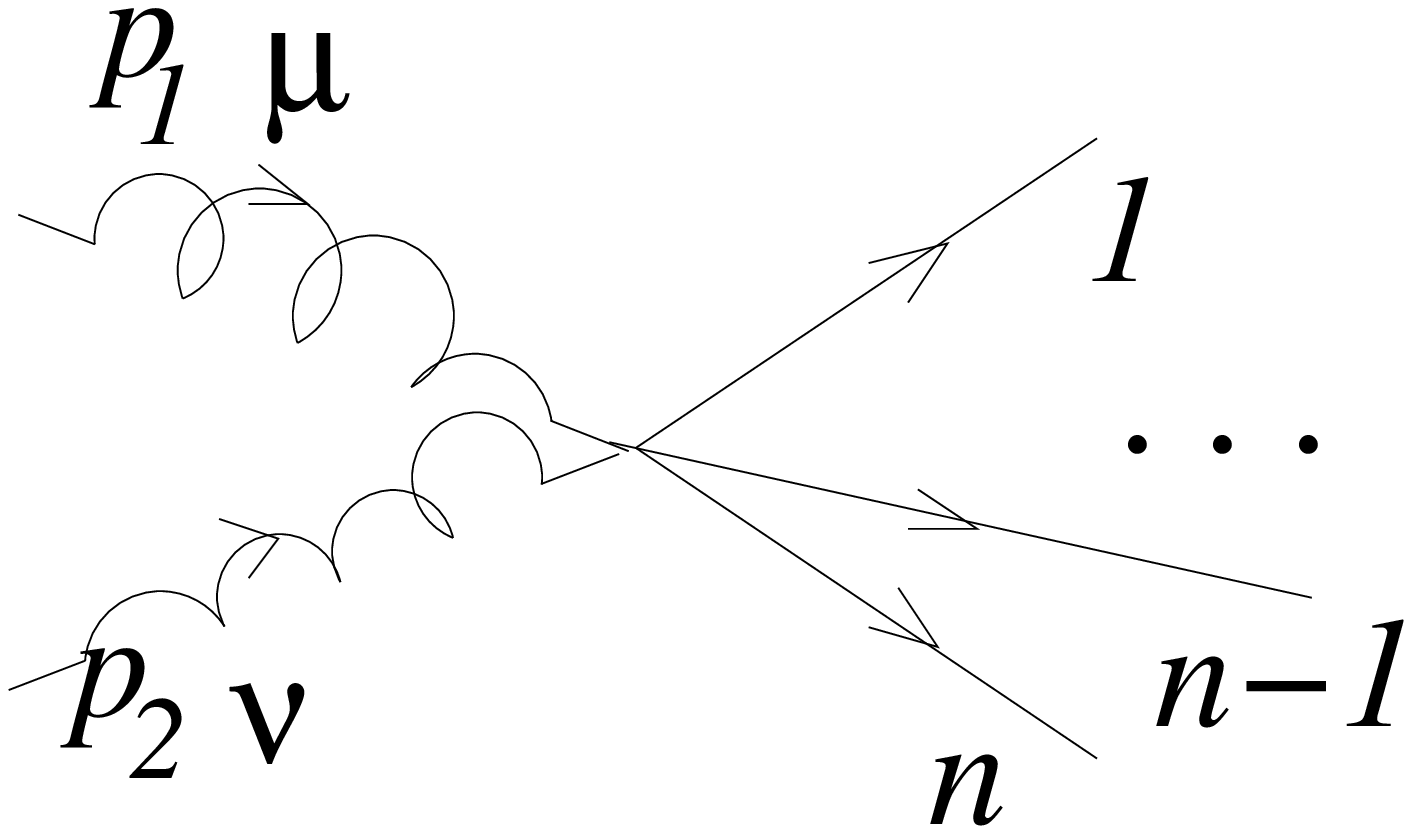,width=4cm}&\mbox{\qquad}&
\epsfig{file=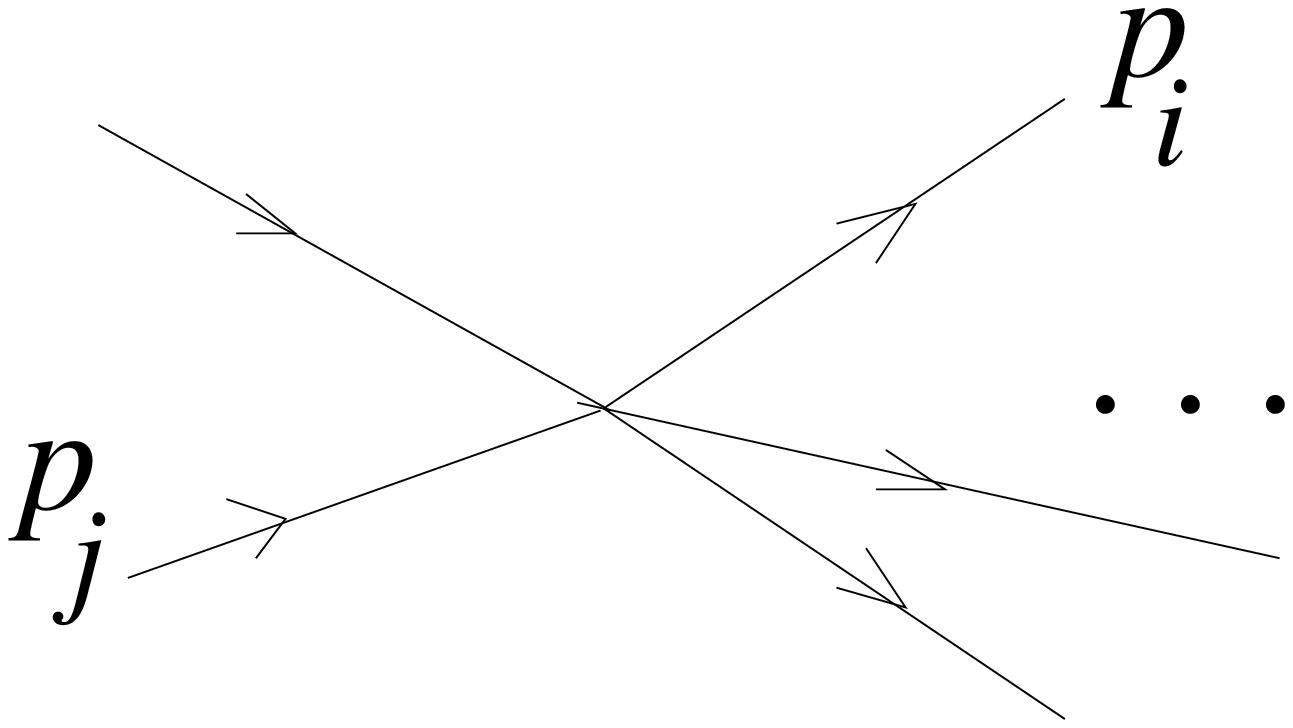,width=4cm}\\
$(n-1)\,(p_1^\nu p_2^\mu-(p_1\cdot p_2)\eta^{\mu\nu})\delta^{ab}$
&&
$-\vac\left\{(n-1)+ \frac{2}{n(n-1)}\sum_{i>j}
\frac{1}{2^{n}m^2}(p_i\cdot p_j)\right\}$\\[0.3cm]
$n=2,3,\ldots$ && $n=3,4,\ldots$\\[0.3cm]
(a)&&(b)
\end{tabular}
\caption{{\sl 
Feynman diagrams of order $g^0$ and given number of
dilatons $n$; (a) gluon-dilaton interaction,  
(b) dilaton self interaction.
}}\label{FIG:FEYNMAN}
\end{figure}

Now, we proceed to the derivation of Feynman rules. In addition to usual
Fyenman diagrams of gluodynamics there are those corresponding to the
dilaton propagator, dilaton self-interactions and dilaton-gluon
interactions. In this work we are interested only in contributions 
of the order of $g^0$. They are shown in \fig{FIG:FEYNMAN}. Other
vertices,
which contain three and four gluons coupled to the dilaton are
supressed by factors $g^1$ and $g^2$ respectively. In our approach the
strong coupling constant is necessarily small since all gluon modes which
enter the lagrangian \eq{LAGR} are hard by construction.  All soft modes,
which carry momenta $p<m$, have been integrated out. 
The dilaton Feynman
propagator can be read from \eq{LAGR}:
\beq\label{PROPAG}
D_d(p)=\frac{m^2}{\vac}\ \frac{i}{p^2-m^2-i\varepsilon}\quad.
\eeq

\begin{figure}[ht]
\begin{center}
\begin{tabular}{cc}
\epsfig{file=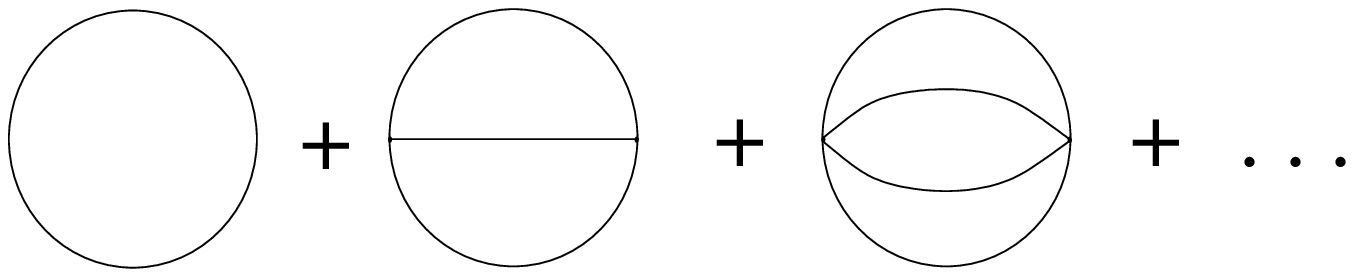,width=7cm,height=1.25cm}
&\epsfig{file=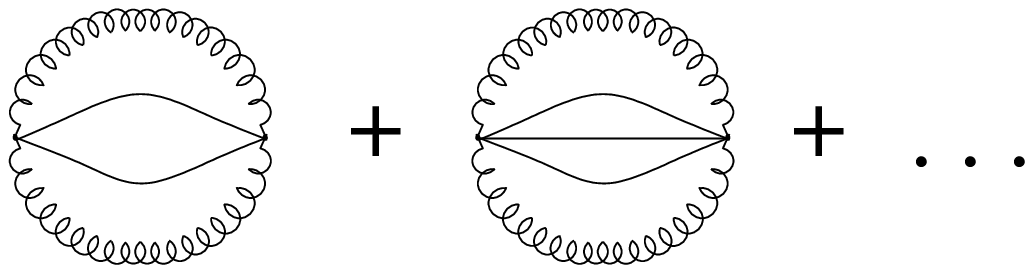,width=5cm,height=1.25cm}\\
(a)&(b)
\end{tabular}
\end{center}
\caption{\sl Some vacuum excitations contributing to
the energy-momentum tensor trace. (a) Dilaton self-interactions;
first diagram is proportional to $\langle\chi^2\rangle$,
(b) dilaton-gluon interactions. 
}\label{FIG:VACUUM}
\end{figure}

To completely specify our effective theory we still have to calculate
the ultra-violet cutoff $M_0$.
Contributions to the vacuum energy density involve loops and therefore
they are divergent in the ultra-violet regime of the theory. Let us
introduce the cut-off $M_0$ which controls
divergences of vacuum excitations contributing to the vacuum expectation
value of the trace of the energy-momentum tensor. The same parameter controls
divergences in the dilaton phase space.  Our requirement that the 
normalization \eq{VAC}, or equivalently \eq{NORMREL}, holds means
that we impose a restriction on the possible values of $M_0$. Consider a 
contribution
stemming from pure dilaton vacuum fluctuations, \fig{FIG:VACUUM}(a).
In the next section we will argue that for the processes we are interested in here 
the phase space is available for
the emission of only two dilatons. In this two-dilaton approximation
\beq
\langle\theta_\mu^\mu\rangle_\mathrm{vac}=
-4\vac\langle e^\chi\rangle_\mathrm{vac}\approx
-4\vac\left(1+\frac{1}{2}\langle\chi^2\rangle_\mathrm{vac}\right)
\quad.
\eeq
Hence, the contribution to the vacuum energy-density is 
\beq\label{CO1}
-2\vac\langle\chi^2\rangle_\mathrm{vac}=
-2\vac\frac{1}{\pi}\int_0^{M_0^2}\frac{dM^2}{M^2}\, \frac{m^2}{\vac}
\int\frac{d^3k}{(2\pi)^3 2\omega}=-\frac{1}{4\pi^3}m^2M_0^2
\quad.
\eeq
Another contribution of the same order in the strong coupling,
$\order{g^0}$, comes from the first diagram shown in
\fig{FIG:VACUUM}(b). We
have
$$
(N_c^2-1)\,\frac{1}{\pi}\int_0^{M_0^2}\frac{dM^2}{M^2}\;
\frac{1}{2!}\int\,\frac{d^3q_1}{(2\pi)^3 2\omega_1}
\int\, \frac{d^3q_2}{(2\pi)^3 2\omega_2}
\Gamma_2\left(q_1^\nu q_2^\mu-(q_1\cdot q_2)\eta^{\mu\nu}\right)^2
\theta(M-\omega_1-\omega_2)
$$
\beq\label{CO2}
=\; \frac{ M_0^8}{280\cdot 48\, 
(2\pi)^6}\,(N_c^2-1)\,\left(\frac{m^2}{\vac}\right)^2 
\quad,
\eeq
where $\Gamma_2$ is a two-dilaton phase-space factor, see \eq{GAMMA}.
Contributions \eq{CO1} and \eq{CO2} cancel out when
\beq\label{CONJ}
M_0^2= 8\pi\, 2^{2/3}\,
105^{1/3}\, (N_c^2-1)^{-1/3}\,\left(\frac{\vac}{m}\right)^{2/3}\quad.
\eeq
The pure gluon vacuum diagrams are proportional to the strong coupling
constant $\alpha_s$ as can be easily seen from \eq{SPUR} and
\eq{BETA}. Inasmuch as the strong coupling constant is small,
those contributions to the vacuum energy density can be neglected. 

\section{Derivation of the soft pomeron}            

\begin{figure}[ht]
\begin{center}
\epsfig{file=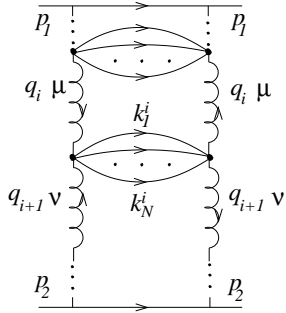,width=4cm}
\caption{\sl One of the ladder diagrams contributing to eq.~\eq{TOT}. 
}\label{FIG:LADDER}
\end{center}
\end{figure}

The leading contribution to the total cross section at high energies stems
from the ladder diagrams \cite{LADD}. Such a diagram is shown in 
\fig{FIG:LADDER}. 
Note that the energy dependence of the Born amplitude is determined by the 
spin of particles 
exchanged in $t$-channel. Thus the leading contribution at high energies
comes from exchange of vector particles -- gluons \cite{LN,GLR}. These gluons prefer to form 
ladders, since
each rung gets enhanced by large logarithm of energy $\ln s\gg 1$.
By virtue of the optical theorem, the total cross section $\sigma_{tot}$
can be calculated as a sum of all such diagrams
\beq\label{TOT}
\sigma_{tot}=\sum_{r\ge 0}\sum_{n\ge 2}\sigma_r^n=
\frac{2}{s}\sum_{r\ge 0}\sum_{n\ge 2}
\mathrm{Im}\,\mathcal{M}_\mathrm{el}(t=0)_r^n\quad,
\eeq 
i.e.\  we sum over
all possible numbers $n\ge 2$ of dilatons produced in $s$-channel in
each rung for a given number $r$ of rungs and then sum over all $r$. 

The most convenient way to calculate the amplitudes associated with ladder
diagrams is to employ the Weizs\"acker-Williams approximation. This means 
replacing almost longitudinally polarized Coulomb gluons exchanged in
$t$-channel by transversely polarized (real) ones. In this approximation the
gluon
propagator reads \cite{GLR}:
\beq\label{WW}
D_g(q^2)\,\approx\,\frac{2}{s}
\frac{q_\bot^\mu q_\bot^\nu}{\alpha_q\beta_q q_\bot^2}\quad,
\eeq
where $\alpha_q$ and $\beta_q$ are Sudakov variables defined by 
\beq
q=\alpha_q\, p_1+\beta_q\, p_2 +q_\bot\quad.
\eeq  
Weizs\"acker-Williams approximation is valid as long as the following
conditions hold
\begin{eqnarray}
&& 1\gg \alpha_{q_1}\gg\ldots\gg \alpha_{q_i}\gg\ldots\gg
\alpha_{q_{r+1}}\quad,\\
&& 1\gg \beta_{q_{r+1}}\gg\ldots\gg \beta_{q_i}\gg\ldots\gg \beta_{q_1}
\quad,\\
&&|q_{i\bot}^2|\ll s\quad.
\end{eqnarray}

In the Born approximation (two gluon exchange) there are no  produced
dilatons. Contribution of a one-rung diagram to the total cross
section  reads
\beq\label{ONER}
\sigma_{r=1}^n=\frac{2}{s}\int\frac{d^4q_1}{(2\pi)^4}
\int\frac{d^4q_2}{(2\pi)^4} \left|
\frac{2g^2(q_1\cdot 
q_2)_\bot \mathrm{Tr\,}(t^at^b)}
{\beta_{q_1}\alpha_{q_2}s}\right|^2
\frac{1}{4N_c^2} (2\pi)^2\delta((p_1-q_1)^2)\delta((p_2+q_2)^2)
\Gamma_n(M)\quad,
\eeq
where $\Gamma_n(M)$ represents contribution of $n$ dilatons to
the phase space and  $M^2=(q_1-q_2)^2$ is the four-momentum squared
available for production in the dilaton vertex. 
The phase space $\Gamma(M)$ has been evaluated in \re{KKL}:
$$
\Gamma_n(M)=
\frac{1}{n!}\left(\frac{m^2}{\vac}\right)^n\, (n-1)^2\,
(2\pi)^4\, \delta^4\left(\sum_{i=1}^nk_i-k\right)
\prod_{i=1}^n \frac{d^3k_i}{(2\pi)^3 2\omega_i}
$$
\beq\label{GAMMA}
=\left(\frac{m^2}{\vac}\right)^n (2\pi)^{4-3n}\,(n-1)^2\,
\frac{(\pi/2)^{n-1}(M^2)^{n-2}}{n!(n-1)!(n-2)!}\quad,
\eeq
where $(n-1)^2$ comes from the vertex shown in \fig{FIG:FEYNMAN}(a).
Using \eq{ONER} and \eq{GAMMA} we derive the contribution of one-ladder
diagram 
\beq\label{R1}
\sum_{n\ge 2}\sigma_{r=1}^n=\sigma_B\frac{\ln s}{(2\pi)^6}\pi^2M_0^2
\int_0^{M_0^2}\, dM^2\, \sum_{n=2}^\infty \Gamma_n(M)=
\sigma_B\Delta_P\ln s\quad,
\eeq
where $\sigma_B\sim\alpha_s$ is a Born amplitude, 
\beq\label{DELTA}
\Delta_P=\frac{a^2}{4\pi}\; _0F_2\left(-;1,3;a\right)
\quad,
\eeq
$_0F_2$ is a generalized hypergeometric function, and the relevant
parameter of our theory for high energy scattering is 
\beq\label{A}
a=\frac{m^2M_0^2}{16\pi^2\, \vac}\quad.
\eeq
Expansion of function $_0F_2$ in powers of parameter $a$ 
\beq\label{EXPAN}
_0F_2\left(-;1,3;a\right)=
1+\frac{a}{3}+\frac{a^2}{48}+\frac{a^3}{2160}+
\order{a^4}
\eeq
is equivalent to expansion in the number of emitted dilatons. We will argue
below that the series \eq{EXPAN} are rapidly convergent  and actually, 
there is not enough phase space for
emission of large number $n$ of dilatons since $\Gamma_n(M)$ is a rapidly
decreasing function of $n$. Thus, we can safely neglect emission of more
than two dilatons.

Now, as we know the contribution of one-rung diagram, calculation of the total
cross section is straightforward. Indeed, the contributions of different rungs
factorize out, leading to the Regge-like behavior of the total cross section
\eq{TOT} 
\beq
\sigma_{tot}=\sigma_B\sum_{r=0}^\infty\frac{1}{r!}\ln^rs\,\Delta_P^r
=\sigma_B\, s^{\Delta_P}\quad.
\eeq
Corrections to the pomeron intercept coming from perturbative QCD are
proportional to the strong coupling constant $g^2$ and neglected in this
approach. Loop corrections to the dilaton propagator give small 
contributions of the order of 
$$
\Delta D_d\,\sim\, \frac{am^2}{M_0^2} \ll 1\quad.
$$
Finally, upon substitution of \eq{A} and \eq{CONJ} into \eq{DELTA} 
we obtain the following result for the soft pomeron intercept
\beq\label{CHTOTO}
\alpha_P=1+\Delta_P=1+\frac{(105/2)^{2/3}}{4\pi^3}\,
(N_c^2-1)^{-2/3}\,
\left(\frac{m^4}{\vac}\right)^{2/3}\quad.
\eeq

To estimate $\vac$ in pure gluodynamics we use the fact that it is related to
the vacuum energy density in QCD $\vac_\mathrm{QCD}$ by
\beq\label{VACREL}
\vac \simeq \frac{11}{9}\,3\,\vac_\mathrm{QCD}\quad.
\eeq
Indeed, as was argued in \re{NSVZ}, 
\beq\label{LKJH}
\langle\alpha_s F_{\mu\nu}^aF^{a\mu\nu}\rangle_\mathrm{glue}\, \approx\, 3\,
\langle\alpha_s F_{\mu\nu}^aF^{a\mu\nu}\rangle_\mathrm{QCD}\quad.
\eeq 
Also the beta function \eq{BETA} in QCD scales with
number of col-ours and flavors as $\beta_\mathrm{QCD}\propto (11N_c-2N_f)/3$ 
whereas  in pure gluodynamics $\beta_\mathrm{glue}\propto
11N_c/3$. Since by \eq{SPUR} $\vac\propto \beta$ the vacuum energy
density in pure gluodynamics gets enhanced by the factor
$11/9$. Together with \eq{LKJH} this yields \eq{VACREL}. Sum rules
analysis of \cite{NSVZ} makes it possible to estimate the QCD vacuum
energy density 
$\vac_\mathrm{QCD}=(0.24\,\mathrm{GeV})^4$. Owing to \eq{VACREL} we
obtain the following estimate:
\beq\label{ESTE}
\vac=0.012 \,\mathrm{GeV}^4\quad.
\eeq

Another parameter of our effective theory is the dilaton mass. Note,
that due to the isospin conservation each dilaton  can produce only even
number of pions in the final state. 
Let us assume that in full QCD with light quarks the dilaton mixes with the 
 broad physical scalar $\sigma$--resonance \cite{SA} at 
\beq\label{MASSA}
m=0.4\div 0.5\, \mathrm{GeV}\quad.
\eeq
Note that the spectral density of the dilaton excitations can be spread over 
several scalar resonances of different mass (see, e.g., \cite{EFK}). However, 
for our purposes it suffices to keep only the lightest 
resonance, which gives the dominant contribution to the intercept because 
of the phase space constraints.

Let us now proceed to the numerical estimates.
In view of  \eq{ESTE} and \eq{MASSA}
Eq.~\eq{CONJ} implies that
$M_0=3.0\div2.8\,\mathrm{GeV}$. 
Substituting these numbers to \eq{CHTOTO} we obtain for the pomeron 
intercept $\alpha_P=1.047\div 1.085$ which is in reasonable  
agreement with the phenomenology of high energy scattering \cite{DL}.
Using these numbers in \eq{EXPAN}, 
we see that the emission of three and more dilatons contributes  
$1- {_0F_2}\left(-;1,3;0\right)/_0F_2\left(-;1,3;a\right)$=$21\div 27\%$
to the final result; this can be used as an estimate of the accuracy of our two-dilaton 
approximation.

\begin{figure}[ht]
\begin{center}
\epsfig{file=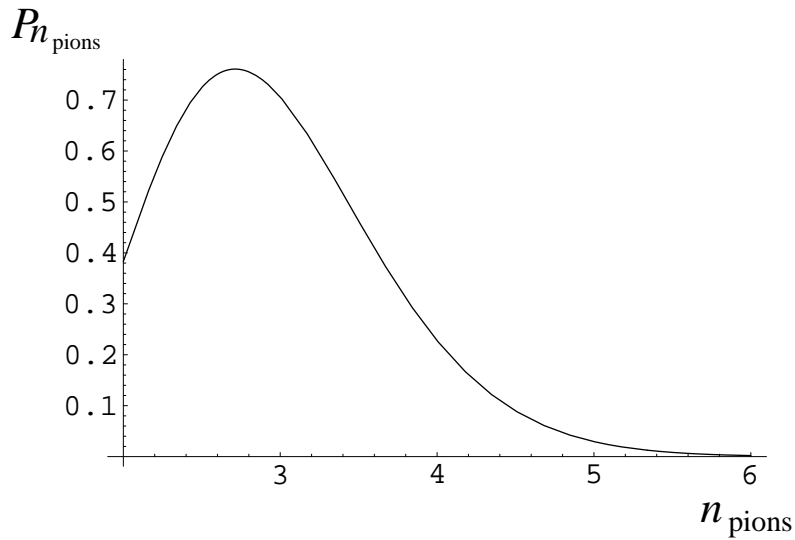,width=11cm}
\end{center}
\caption{\sl Number distribution of charged pions produced by dilaton
mechanism for $a=0.9$ at hadron-hadron collisions.
 }\label{FIG:SPECTR}
\end{figure}

We can calculate the inclusive spectrum of dilaton production by observing that
Eq.~\eq{GAMMA} is nothing but the number distribution of the dilatons
produced per unit of rapidity. Indeed, each rung of the ladder
diagram corresponds to one unit in rapidity (multiplied by $\Delta_P$) in
the leading logarithmic
approximation. Define the probability $P_n(a)$ to find $n$ dilatons in
the final state
\beq\label{PROBAB} 
P_n(a)=\frac{\int_0^{M_0^2}dM^2\;\Gamma_n(M)}
{\sum_{n=2}^{\infty}\int_0^{M_0^2}dM^2\;\Gamma_n(M)}\quad.
\eeq
Then the average number of dilatons produced per unit of rapidity is
given by
\beq\label{DESTRIB}
\langle \frac{dN_\chi}{d(\Delta_P\, y)}\rangle = \sum_{n=2}^{\infty}n\,
P_n(a)=
\frac{2\; _0F_2\left(-;1,2;a\right)}{ _0F_2\left(-;1,3;a\right)}\quad.
\eeq
From \eq{A} we get an estimate of the dilaton phase space parameter
$a=0.768\div 1.033$. Substituting this value into \eq{DESTRIB} one arrives at 
$\langle dN_\chi/dy\rangle=0.1\div 0.2$. Since we have identified the
dilaton with the 
physical $\sigma$-resonance, we have to assume that it decays 
mostly in two pions. Charged pions give
approximately 2/3 of the total number of pions. Therefore, the average multiplicity of
charged pions per unit of rapidity is estimated as
\beq
\langle \frac{dN_\pi^\mathrm{ch}}{dy}\rangle\, =\, \frac{2}{3}\,
2\, \langle \frac{dN_\chi}{dy}\rangle\, =\,  0.14\div 0.26 \quad.
\eeq
This agrees with the result obtained in\cite{KHALE}: the soft pomeron
is responsible for less then 10\% of the observed multiplicity. 
In \fig{FIG:SPECTR} we show the number distribution \eq{PROBAB} 
of produced charged pions.

\section{Summary and discussion}

Summarizing, we have started with the derivation of the QCD effective lagrangian \eq{LAGR}
which incorporates scale anomaly\cite{SA,MISHI} and represents propagation 
and interaction of gluons and dilatons. The
dilaton field saturating the sum rules \eq{LET} describes 
the QCD vacuum structure.  The pure dilaton part of lagrangian \eq{LAGR} (first and second terms) is 
a low energy effective gluodynamics obtained by integrating
out  low energy modes with momenta $q$ such that $q\le
m\approx 0.4\div 0.5$ GeV. The third term of
\eq{LAGR} represents pure gluodynamics and the interaction of gluon field with
the dilaton one. It is important as far as momentum transfer does not
exceed $M_0$. It has been already pointed out in
\re{NSVZ,FK,KHALE} that the scale $M_0$, determined by the semi-classical fluctuations of 
the gluon fields in
vacuum is quite large, $M_0\gg\Lambda$. 

Normalizing the vacuum energy density by
\eq{VAC}  we obtained an  expression \eq{CONJ} for the ultra-violet
cutoff $M_0$ of the theory in terms of vacuum energy density $\vac$
and the dilaton mass $m$. Divergent integrals that we have encountered
in Eqs.~\eq{CO1},\eq{CO2},\eq{R1} should not
surprise the reader since the dilaton acts as the gravitational field. 
Indeed, quantum gravity
is a non-renormalizable theory  since the coupling of gravity to matter
fields $G$ has canonical dimension (mass)$^{-2}$. In technical terms, each
emitted dilaton  contributes a factor of  $m^2/\vac$ coming from the
normalization of the dilaton propagator which must be compensated for
by  $M_0^2$ to maintain the scattering amplitude dimensionless. 

In the framework of this effective theory  we addressed the 
problem of soft pomeron. We calculated the high energy scattering
amplitude in the leading logarithmic approximation by summing the
ladder diagrams, see \fig{FIG:LADDER}. Generally, any number of
dilatons  can be emitted in the $t$-channel in each ladder rung, but 
actually most of the phase space is occupied by only two of them. Owing to
the isospin conservation the dilaton in the final state decays into an 
even number of pions. 
By identifying the dilaton with the physical $\sigma$ resonance, 
we conclude that it should decay mostly in two pions. 
We estimate the pomeron intercept 
$\alpha_P=1.047\div 1.085$. This value agrees well with the analysis of 
the data on high energy scattering \cite{DL}. 

We also estimate the final state multiplicity per unit
of rapidity $\langle \frac{dN_\chi}{dy}\rangle \simeq 0.2$. 
The low particle multiplicity resulting from the soft pomeron exchange 
has been noted already in \cite{KHALE}. It implies that the bulk of
particles is produced by another mechanism.  
Such a mechanism must involve multiple production of particles per unit of 
rapidity and perhaps closely related to high parton density regime of QCD
\cite{GLR,MV}.

The reader familiar with the previous work of two of us, \re{KHALE}, might
be confused by the fact that there the opposite to \eq{CHTOTO} dependence 
of $\Delta$ on the vacuum energy density $\vac$ and the scalar particle
(glueball) mass $m_R$ has been found. The reason is simple: a glueball
mediates a short-range strong interaction, so the cross section for its
production
is proportional to $m_R^{-4}$. On the other hand, gravitation is a
long-range interaction, the strength of which is proportional to masses  
of interacting particles (dilatons). In other words, the effective lagrangian 
\eq{LAGR} is constructed to obey scale invariance on the classical level. 
This requirement translates into the absence of the coupling of the dilaton to 
two gluons, because a term linear in the dilaton field $\chi$ in \eq{LAGR} would 
induce a non-zero vacuum expectation value $\langle \chi \rangle$ which 
would violate the scale invariance already on the classical level. 
(Even though the field $\chi$ is dimensionless, its couplings are not, and 
so its v.e.v. would violate scale invariance of the vacuum state). 
This is why the coupling of the scalar glueball to two gluons, used in \cite{KHALE}, 
does not appear in the approach of this paper, which explains the difference in 
the dependence of the final results on the scalar particle mass.        
 
The reader may also wonder about the meaning of coupling QCD to 
gravity. 
One way of interpreting this 
is to say that coupling to gravity is a purely mathematical trick
which makes it possible to take all symmetries into account in the most
elegant way \cite{MISHI}. 

However, there might be another interpretation
which goes beyond the standard model. We treated the gravitational 
constant $G$ as a free
parameter. We can find its numerical value using \eq{MASS}: 
$G=0.2$~GeV$^{-2}$. One might thus assume that  
gravitational interactions become strong already at the scale
$G^{-1/2}\sim$2.2~GeV. In this case cutting divergent integrals
associated with dilaton fields at the scale of  $M_0$ means that the discussed
theory \eq{LAGR} is an effective one. It implies that the full
renormalizable theory must be recovered at some higher energies.
In this sense our scale $M_0$ can be derived from some
fundamental scale $M^*$. In recent years
we have learnt that the existence of strong gravity regime at experimentally
accessible energies can be realized in theoretical models if one 
assumes that our world has dimension larger then four. The Standard
Model is confined to the four-dimensional manifold, while gravity
propagates in all dimensions. One further assumes that there is no
specific gravity scale, but rather a unique fundamental scale
for the Standard Model and gravity 
around the electroweak scale $M^*\sim 1$~TeV \cite{DIM}. 
Thus the observed Plank scale is  small since the size of extra
dimensions is large (up to mm). This does not seem to contradict
experimental data since the gravitational potential has
been tested at distances down to a cm.
Existence of strong gravitational interactions at TeV scale has a
number of remarkable experimental signatures \cite{DIM,NUS,GIDD}. On  
the contrary, one may assume that the scale of gravity is different
from the Standard Model one, but the gravitational lagrangian involves
higher derivatives of the metric tensor \cite{DVALI}. In this approach the
quantum gravity scale can be as low as $10^{-3}$~eV. 
However, we explicitly neglected all higher order derivatives of
metric in \eq{LAGR}. 
Even if the quantum gravity indeed becomes strong at 
$M^*\ll M_\mathrm{Pl}$  the probability that a dilaton escapes
 into higher dimensions is
suppressed by factor of order of $\order{s/M^{*2}}$, which means
that our four-dimensional effective lagrangian \eq{LAGR} is a good
approximation at energies $\sqrt{s}\ll M^* $. So, at these energies gravity
can be considered as a background classical field, since quantum
corrections are small. 
It would be interesting to address the question of what is the mechanism
by which the full theory at the fundamental scale yields the
conformally flat GeV-scale gravitational effects. However, we leave
this problem aside.

Whatever the underlying mechanism is, the broken scale invariance   
can be encoded in the effective low--energy theory of gluodynamics in the form of 
the lagrangian \eq{LAGR}. 
We hope that it 
 provides a useful tool for a
systematic approach to non-perturbative processes at high energies.


\vskip1.cm
{\large\bf Acknowledgements}
\vskip0.3cm
We are grateful to J.~Ellis, E.~Gotsman, A.B.~Kaidalov, D.B.~Kaplan, 
Yu.~Kovchegov, U.~Maor,  
L.~McLerran, M.~Pospelov, E.~Shuryak, C-I~Tan and I.~Zahed
for illuminating discussions on the subject and helpful comments. 
The work of D.K. was supported by the U.S. Department of Energy under Contract 
No. DE-AC02-98CH10886. 
This research was supported in part by the  BSF grant \#98000276, by the 
GIF grant \# I-620-22.14/1999 and by Israeli Science Foundation, founded 
by the Israeli Academy of Science and Humanity. The work of K. T. was 
sponsored in part by the U.S. Department of Energy under Grant
No. DE-FG03-00ER41132. The research of D.K. and K.T. 
was supported in part by the National Science
Foundation under Grant No. PHY94-07194.

 \vskip1.cm


\end{document}